\documentclass[aps,10pt,floatfix,twocolumn,showpacs]{revtex4-1} 
\usepackage{color,soul}
\usepackage{graphicx}
\usepackage{siunitx}
\usepackage[breaklinks=true,colorlinks=true,linkcolor=blue,urlcolor=blue,citecolor=blue]{hyperref}

\usepackage{dblfloatfix}
\usepackage{amsmath, amsthm, amssymb}
\usepackage{amsthm}
\usepackage{multirow}
\usepackage[normalem]{ulem}
\usepackage{soul}
\usepackage{makecell}
\usepackage[table]{xcolor}
\definecolor{Gray}{gray}{0.85}
\definecolor{LightCyan}{rgb}{0.88, 1, 1}
\definecolor{Apricot}{rgb}{0.98, 0.81, 0.69}
\usepackage{threeparttable}
\newcommand{\be}{\begin{equation}}
\newcommand{\ee}{\end{equation}}
\newcommand{\bea}{\begin{eqnarray}}
\newcommand{\eea}{\end{eqnarray}}

\usepackage[rightcaption,]{sidecap}
\sidecaptionvpos{figure}{c}
\usepackage[utf8]{inputenc}
\usepackage{amsmath}
\usepackage{amssymb}
\usepackage{graphicx}
\usepackage{hyperref}
\usepackage[export]{adjustbox}
\usepackage[update]{epstopdf}
\usepackage{multirow}
\usepackage{placeins}

\begin{document}

\title{Neutral polymer conformations with attractive bridging crowder interactions: role of crowder size}
\author{Hitesh Garg}
\email{hiteshgarg@imsc.res.in}
\affiliation{The Institute of Mathematical Sciences, C.I.T. Campus,
Taramani, Chennai 600113, India}
\affiliation{Homi Bhabha National Institute, Training School Complex, Anushakti Nagar, Mumbai, 400094, India}

\author{Satyavani Vemparala}
\email{vani@imsc.res.in}
\affiliation{The Institute of Mathematical Sciences, C.I.T. Campus,
Taramani, Chennai 600113, India}
\affiliation{Homi Bhabha National Institute, Training School Complex, Anushakti Nagar, Mumbai, 400094, India}

\date{\today}
\begin{abstract}
Extensive molecular dynamics simulations were conducted to explore the conformational phase diagram of a neutral polymer in the presence of attractive crowders of varying sizes. For weakly self-attractive polymers, larger crowders induce a transition from an extended state to a bridging-crowder-induced collapse (CB phase) at lower polymer-crowder interaction strengths. Similarly, for strongly self-attractive polymers, a transition from a crowder-excluded collapse (CI phase) to the CB phase occurs with larger crowders and stronger polymer-crowder attraction. The simulations reveal that the collapsed states in both weakly and strongly self-attractive polymers are identical when small bridging crowders are involved. Near the transition threshold for weakly self-attractive polymers, increasing crowder size leads to complex transitions, including collapse, extension, and re-collapse. Additionally, a confined collapse (CC phase) is observed, where the polymer is confined to interstitial spaces among large crowders. Strongly self-attractive polymers show similar transitions, but only with smaller crowders, as larger crowders suppress these changes. These findings underscore the crucial influence of crowder size on the conformational behavior of neutral polymers, especially under varying degrees of polymer-crowder interaction.
\end{abstract}

\maketitle
\section{Introduction}
The influence of crowders on polymer conformation is significant in biological systems, where many biopolymers exist in crowded environments~\cite{zimmerman1991estimation,ellis2003join,ellis2001macromolecular,zimmerman1993macromolecular,zhou2008macromolecular,ellis2001macromolecular1,foffi2013macromolecular}. These environments contain crowders with varying sizes and shapes. The size of these crowders affects several polymer properties, including compaction, diffusion, hydration structure, polymer looping kinetics, translocation, and the glassy behavior of bacterial cytoplasm~\cite{kang2015effects,tan2021effects,shin2015kinetics,wang2017influence,parry2014bacterial}. Understanding how crowders influence synthetic and biological macromolecules is crucial, given the diversity of interactions that arise in crowded environments~\cite{nayar2020small,zangi2009urea,mardoum2018crowding,nakano2017model,shin2015kinetics}. In polymer looping, for example, small crowders increase environmental viscosity, slowing down chain dynamics and reducing the rates of looping and unlooping~\cite{shin2015kinetics,zhang2021polymer,zhang2021comparative}. Conversely, larger crowders enhance excluded volume effects, bringing polymer segments closer together and potentially increasing the likelihood of loop formation. Previous studies have also shown that crowder size is key in determining the onset of collapsed conformations in neutral polymers under high crowder density and confined conditions~\cite{zhou2008macromolecular,asakura1958interaction,bhat1992steric}. Specifically, smaller crowders induce greater polymer compaction at equal density, suggesting that reduced crowder size can drive transitions from extended to collapsed polymer conformations. Traditionally, crowders in polymer science are modeled as steric or volume-excluded entities, a concept central to understanding depletion-induced polymer collapse, which is largely entropic~\cite{minton1981excluded,zhou2008macromolecular,asakura1958interaction,bhat1992steric,minton1981effect}. Interestingly, both very large and very small crowders can trigger polymer collapse transitions~\cite{kang2015effects}. Large crowders induce collapse primarily via depletion forces: as crowder size increases beyond a critical value, these particles are entropically driven to vacate the polymer’s vicinity, thereby stabilizing the collapsed conformation by maximizing available volume for crowders~\cite{zhou2008macromolecular,asakura1958interaction,bhat1992steric}.

Biological studies highlight that crowded cellular environments often feature a combination of non-specific soft attractive interactions and hard volume exclusion, both of which critically affect biomacromolecular interactions and conformations~\cite{sarkar2013soft,sagle2009investigating,zangi2009urea,street2006molecular}. Soft attractive interactions typically destabilize macromolecules and promote unfolding~\cite{miklos2010volume,lim2009urea,sagle2009investigating}. However, recent evidence suggests that, under certain conditions, attractive interactions between polymers and crowders can drive coil-globule transitions, stabilizing collapsed or folded conformations~\cite{antypov2008computer,heyda2013rationalizing,rodriguez2015mechanism,sagle2009investigating,huang2021chain,ryu2021bridging,brackley2020polymer,brackley2013nonspecific,barbieri2012complexity,garg2023conformational}. This phenomenon occurs when polymer-crowder attractions exceed a critical threshold, allowing crowder particles to act as 'bridging interactions' that connect distant monomers along the polymer chain. These bridging interactions enhance intra-polymer attractions, leading to polymer collapse. Unlike purely repulsive, athermal interactions, which primarily drive polymers through entropic mechanisms, tunable soft attractive interactions expand the range of possible macromolecular conformations~\cite{jiao2010attractive,kim2013crowding}. Recent studies~\cite{heyda2013rationalizing,rodriguez2015mechanism,sagle2009investigating,huang2021chain,tripathi2023conformational} have shown that attractive crowders can induce a polymer collapse transition when the polymer-crowder attraction exceeds a certain threshold. In these cases, crowders facilitate collapse by acting as 'bridges' between distant monomers, particularly when the monomers themselves exhibit repulsive interactions. However, the exact mechanism by which crowder size influences this collapse, especially the critical point of monomer-crowder interaction, remains unresolved. While the effects of repulsive crowder size on polymer compaction are well documented~\cite{liu2020non,kang2015effects,sharp2015analysis,kim2015polymer,chen2015polymer}, the role of attractive crowder size in determining polymer conformation is still an open area for further investigation.

In our study, we investigate two primary aspects: the conformational phase diagram of polymers, influenced by the size of crowder particles and the strength of polymer-crowder attractive interactions, and the impact of crowder size on effective bridging interactions. We first demonstrate that attractive crowders can induce polymer collapse at much lower densities than those required for depletion-mediated collapse, particularly when crowder sizes are comparable to the monomer size. This contrasts with depletion-type collapse, where very small crowders are most effective in driving the transition. By examining weakly and strongly self-interacting polymers, we identified three distinct phases: a collapsed phase (CI), characterized by strong intra-polymer attraction without crowders inside the structure; an extended polymer phase (E); and a bridging-induced collapsed phase (CB), where a substantial number of crowders are incorporated within the collapsed polymer. Notably, the CB phases in weakly and strongly self-interacting polymers are identical. The phase boundaries were determined by analyzing the polymer’s radius of gyration, the number of bridging crowders, and the density of crowders within the collapsed structures. Our findings also reveal two distinct collapse mechanisms for weakly self-interacting polymers: (1) Bridging Collapse (CB phase), where small crowders bridge polymer monomers within the structure, and (2) Confining Collapse (CC phase), where larger crowders confine the polymer by creating interstitial spaces, and the polymer collapses by occupying these spaces. These results highlight how crowder size and polymer-crowder interactions collaboratively shape the conformational landscape of neutral polymers in crowded environments.

\section{Model and Methods}\label{Sec-2}

We consider a coarse-grained bead-spring model for a linear homopolymer consisting of $N_m$ identical monomers in a volume $V$ in the presence of $N_c$ crowder particles. The adjacent monomers are connected by harmonic springs. A pair of non-bonded particles at distance $r$ interact through the Lennard-Jones (LJ) potential:
\be
V_{LJ} (r)= 4\epsilon_{ij}\left[ \left(\frac{\sigma_{ij}}{r_{ij}} \right)^{12}-\left(\frac{\sigma_{ij}}{r_{ij}} \right)^6 \right],
\ee
where $i,j$ takes on values $m,c$ depending on whether the pair is monomer-monomer, crowder-crowder, or monomer-crowder. In the simulations, we truncate the LJ potential at $r_c=3 \sigma_{ij}$.

The monomers of the linear polymer are connected via harmonic springs: 
\begin{equation}
V_{bond}(r)=\frac{1}{2} k (r_b-b)^2,
\end{equation}
where we set $b=1.12 \sigma_{m}$, $r_b$ is the separation between the bonded particles, and $k$ is the stiffness constant. We choose $k=\frac{500 \epsilon_{mm}}{\sigma_{m}^2}$.

We simulate polymers of length $N_m=100$, unless otherwise stated. We simulate systems with $\epsilon_{mm}=0.1, 1.0$, representing weakly- and strongly self-interacting polymers. For all the simulations, we take the crowder-crowder interaction $\epsilon_{cc}=1.0$. For each value of $\epsilon_{mm}$ and $\sigma_{c}/\sigma_{m}$, we simulate systems with $\epsilon_{mc}$ varying from $0.1$ to $4.0$. The crowder volume fraction is ($V$ is the volume of the simulation box, which is $(40\sigma_m)^3$):
\be
\phi_c=\frac{N_c}{V} \frac{4}{3} \pi \left( \frac{\sigma_{c}}{2}\right)^3.
\ee

For all simulations, we fix the crowder volume fraction at $\phi_c = 0.025$, regardless of crowder size. This low volume fraction ensures that the observed transitions are driven by the interplay between crowder size and polymer-crowder attractive interactions, rather than depletion-like interactions, which typically occur at higher volume fractions or crowder densities. This value is well below the threshold required to induce collapse via depletion-mediated interactions for all crowder sizes considered (as will be discussed in the next section). The number of crowder particles, $N_c$, corresponding to different crowder sizes is provided in Table~\ref{table1}.
\begin{table}		
\caption{\label{table1} Number of crowders $N_c$ in the different simulations, chain length is $N_m=100$ unless stated otherwise. The number of crowders correspond to a crowder volume fraction of $\phi_c$=0.025.}
\begin{ruledtabular}
		\begin{tabular}{lll}
$\sigma_{c}/\sigma_{m}$  &$N_c$ \\
		\hline
$0.40$ & 47,700 \\
$0.50$ & 24,456  \\
$0.60$ & 14,000  \\
$0.75$ & 7,400  \\
$0.90$ & 4,190  \\
$1.00$ & 3125  \\
$1.50$ & 925 \\
$2.00$ &390  \\
$2.50$ & 200  \\
$3.00$ & 115  \\
$4.00$ & 48  \\
$5.00$ & 25  \\
\end{tabular}
\end{ruledtabular}
\end{table}

The equations of motion are integrated using the MD LAMMPS software package~\cite{plimpton1995fast}, and visualization of images and trajectories is performed with the Visual Molecular Dynamics (VMD) package~\cite{HUMP96}. The time step is set to $\delta t = 0.001\tau$, where $\tau = \sigma_{m}\sqrt{\frac{M_m}{\epsilon_{m}}}$, with $M_m$, $\sigma_m$, and $\epsilon_{mm}$ representing the units of mass, length, and energy, respectively, defined in terms of the monomers. To achieve the desired density, the initial configuration is equilibrated in the NPT ensemble for $10^7$ time steps. Each subsequent simulation is run for $2 \times 10^7$ steps using the velocity-Verlet algorithm under constant volume and temperature conditions, with the temperature fixed at $T=1.0$ using a Nos'{e}-Hoover thermostat. For analysis, block averaging is performed over the last $n=900$ frames of each simulation, with each block consisting of 100 frames, resulting in a total of $n_b=9$ blocks. The standard deviation of the blocks relative to the mean is computed and divided by $\sqrt{n_b}$ to estimate the error for each parameter. This methodology ensures statistical reliability in the reported averages while accounting for any fluctuations in the system.

\section{\label{sec:results} Results}

\subsection{\label{sec:lowEmm} Extended to collapse transition via bridging crowders}
\begin{figure}
\includegraphics[width=0.9\columnwidth]{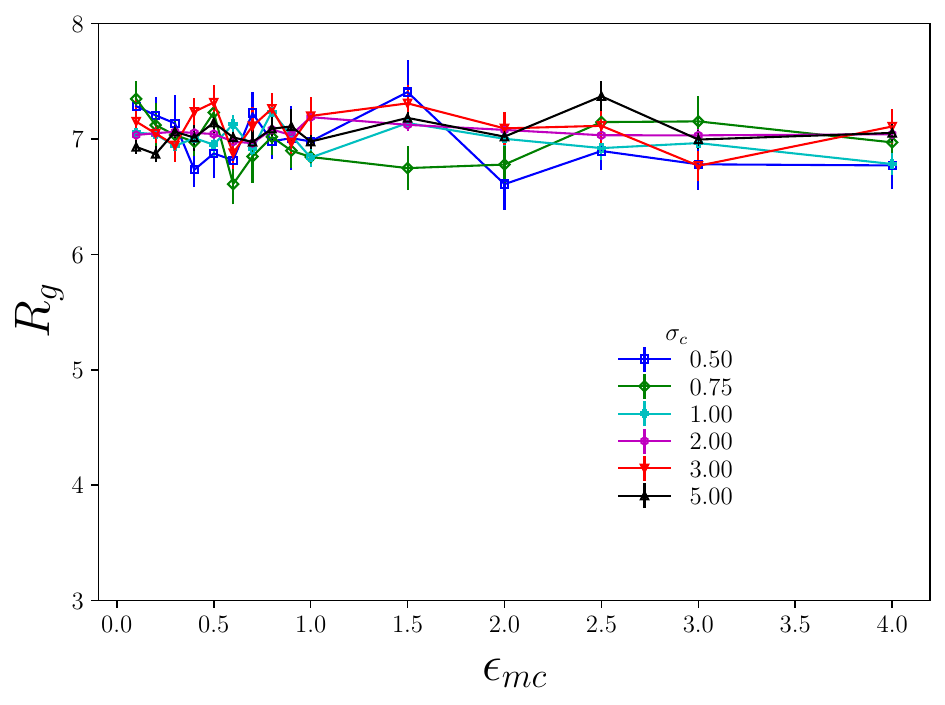}
 \caption{Radius of gyration ($R_g$) of the polymer as a function of crowder size at a crowder density of $\phi_c=0.025$, for purely repulsive polymer-crowder interactions of varying strengths.}
\label{rg-wca}
\end{figure}
In this section, we examine the role of crowder size in driving the transition from extended to collapsed polymer conformations by considering a very weakly self-attractive polymer ($\epsilon_{mm}=0.1$) while varying the polymer-crowder interaction strength ($\epsilon_{mc}$). The crowder-crowder interactions are kept attractive and constant throughout the simulations. The conformations of the polymer are characterized by its radius of gyration, $R_g$:
\begin{equation}
    R_g^2=\frac{1}{N_m}\sum_{i=1}^{N_m}(\textbf{r}_i-\textbf{r}_{\mathrm{cm}})^2,
    \label{eq:rg}
\end{equation}
Notably, at the crowder densities used in this study ($\phi_c=0.025$), purely repulsive interactions do not induce polymer collapse for any crowder size, regardless of interaction strength (see Figure \ref{rg-wca}). Furthermore, previous studies have shown that depletion-induced collapse with smaller crowders only occurs at higher crowder densities ($\phi_c > 0.3$)~\cite{kang2015effects}. In contrast, we demonstrate that in the presence of soft attractive interactions, transitions from extended to collapsed conformations occur across the same range of crowder sizes, even at this lower crowder density. We systematically explore the conformations of a neutral polymer in the $\epsilon_{mc}$--$\sigma_{c}$ plane. For all simulations in this section, the crowder volume fraction is set to $\phi_c=0.025$, and the crowder-crowder interaction strength is fixed at $\epsilon_{cc}=1.0$. These results highlight that attractive polymer-crowder interactions can induce polymer collapse even at low crowder densities, where depletion forces are insufficient.
\begin{figure}
\includegraphics[width=\columnwidth]{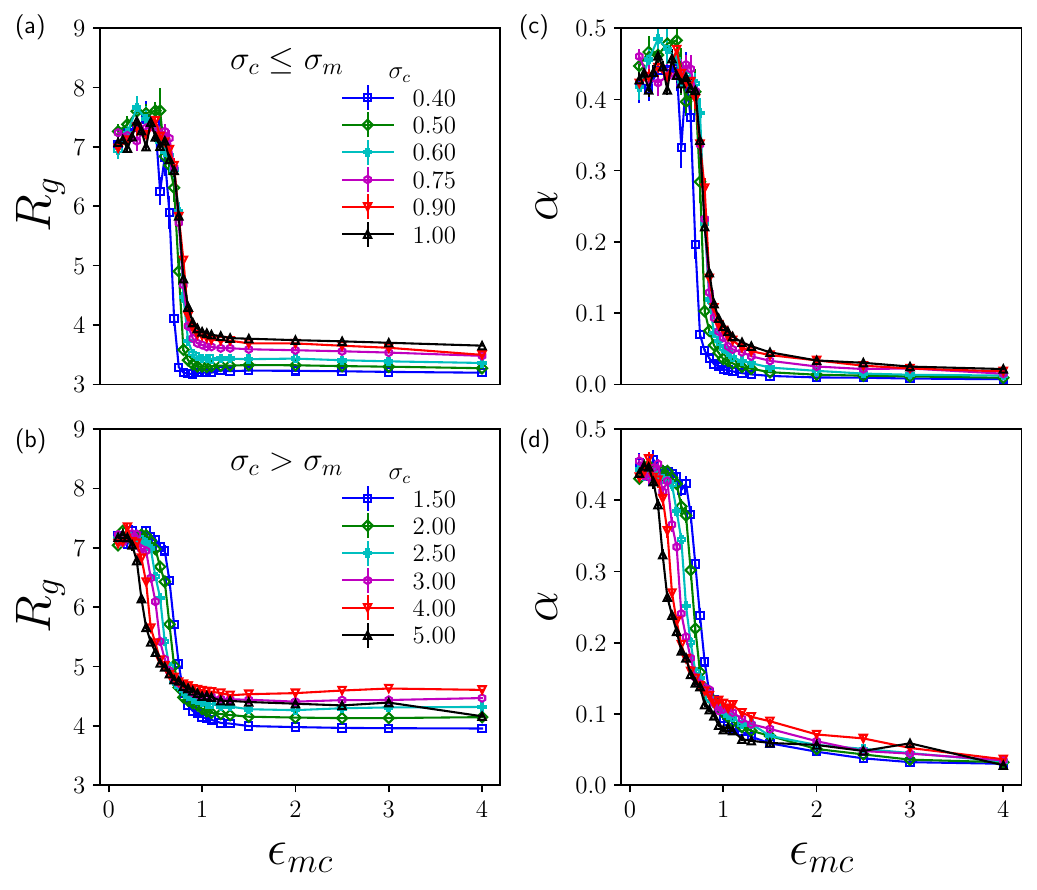}
 \caption{Variation of radius of gyration, $R_g$ (a, b), and asphericity, $\alpha$ (c, d), as a function of polymer-crowder attractive interaction strength, $\epsilon_{mc}$, for $\sigma_c \leq \sigma_m$ and $\sigma_c > \sigma_m$. The data are for weakly self-interacting polymer, $\epsilon_{mm}=0.1$.}
\label{rg-epsilonlow}
\end{figure}

\begin{figure}
\includegraphics[width=0.8\columnwidth]{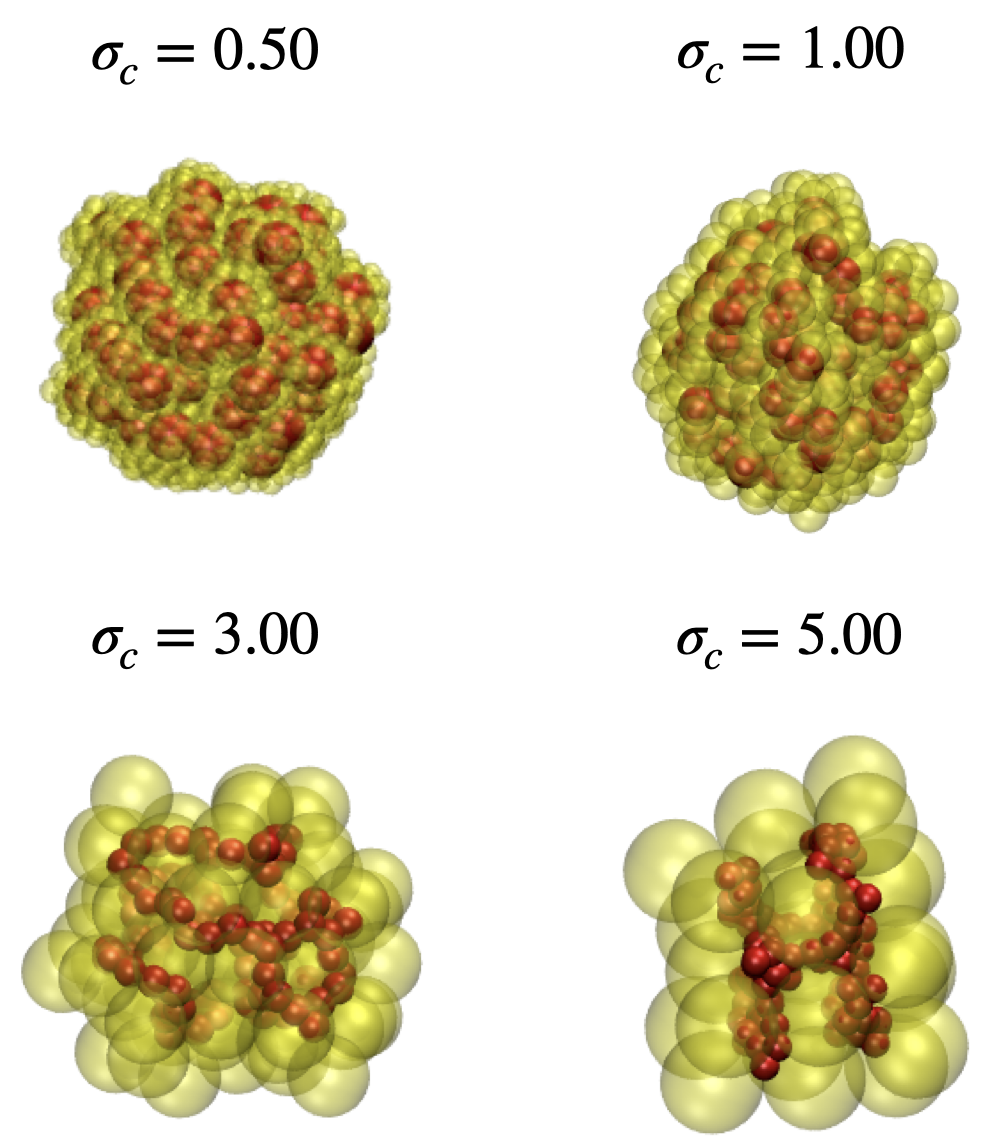}
 \caption{Polymer conformations (in red) as a function of $\sigma_c$ for highest $\epsilon_{mc}=4.0$ considered. The crowders, within $1.5\sigma_{mc}$, are shown in transparent representation in yellow colour.}
\label{varySigma}
\end{figure}

Figures \ref{rg-epsilonlow}(a) and (b) show the variation in the radius of gyration ($R_g$) as a function of the monomer-crowder interaction parameter, $\epsilon_{mc}$, for cases where $\sigma_c \leq \sigma_m$ and $\sigma_c > \sigma_m$, respectively. Across all crowder sizes, the polymer transitions from an extended to a collapsed state as $\epsilon_{mc}$ increases from a low value of 0.1. This transition is reflected in the asphericity values shown in Figures \ref{rg-epsilonlow}(c, d). Notably, the polymer size differs significantly between small and large crowders at high $\epsilon_{mc}$ values, as seen in Supplementary Figure S1.  These discrepancies primarily arise from whether the crowders are located within and between the monomers (in the case of small crowders) or whether the polymer occupies the interstitial spaces between larger crowders, as shown in Figure \ref{varySigma}. As the crowder size increases, a transition occurs from small crowders bridging the polymer to the polymer itself bridging the larger crowders. In the latter case, the collapsed polymer occupies the interstitial spaces between the large crowders and adopts finger-like structures. This transition from small to large crowder effects is further illustrated by the pair correlation functions $g_{mm}(r)$ and $g_{mc}(r)$ for the smallest and largest crowder sizes at high $\epsilon_{mc}$ in the collapsed phase, as shown in Supplementary Figure S2. For $\sigma_c=0.4$, crowders are found closer to monomers than to other monomers, indicating that the small crowders bridge the monomers within the collapsed phase. Conversely, for $\sigma_c=5.0$, monomers are more likely to cluster near each other, while the polymer occupies the spaces between the large crowders. 

\begin{figure}
\includegraphics[width=\columnwidth]{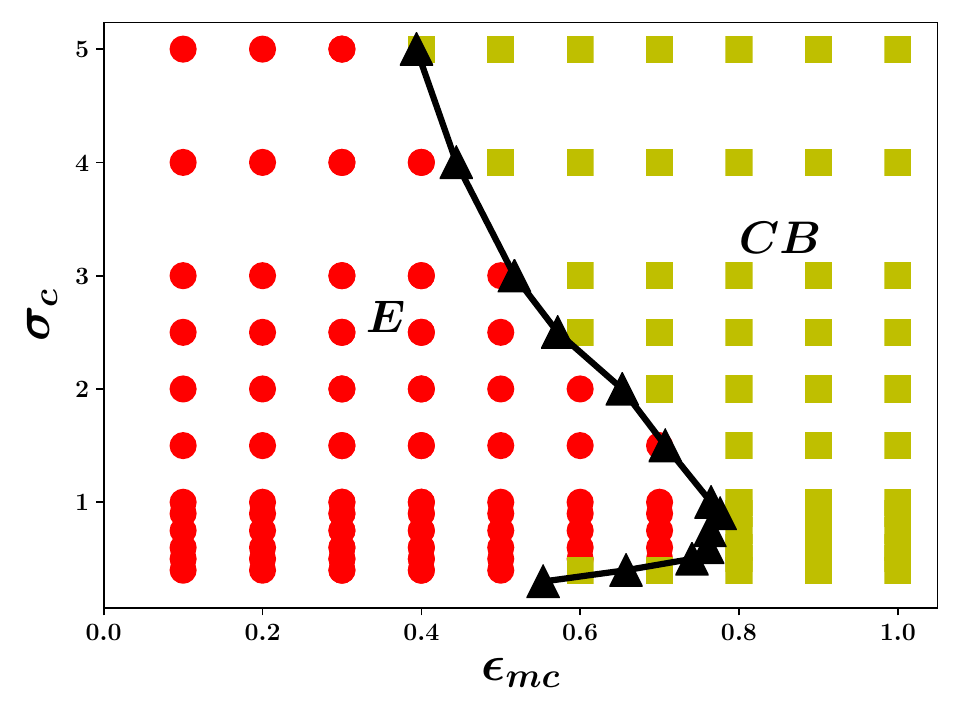}
 \caption{Phase diagram for polymer conformation in $\sigma_{c}$-$\epsilon_{mc}$ plane for weakly self-attractive neutral polymer, $\epsilon_{mm}=0.1$, obtained from the analysis of the chain structure in terms or radius of gyration $R_g$. The simulated systems corresponding to extended phase (E) and bridging crowder induced collapsed phase (CB) are colored in red and yellow respectively. The phase boundary is shown in black.}
\label{phasediagramemmlow}
\end{figure}

The analysis of the radius of gyration, $R_g$, (as shown in Figures \ref{rg-epsilonlow}(a) and (b)) clearly reveals two distinct polymer phases: (1) an extended phase ($E$), characterized by weak intrapolymer interactions and low polymer-crowder interaction values ($\epsilon_{mc}$) across all crowder sizes, and (2) a collapsed phase ($CB$), driven by bridging attractive interactions facilitated by crowders at higher $\epsilon_{mc}$ values. To delineate these phases within the $\epsilon_{mc}-\sigma_c$ phase space and accurately identify the phase boundary, we focus on determining the transition points between these phases. These transition points ($\epsilon_{mc}^*$) are identified at values of $\sigma_c$ and $\epsilon_{mc}$ where the rate of change in the radius of gyration ($dR_g/d\epsilon_{mc}$) is maximal. To precisely determine these gradients, we apply a fitting procedure using a hyperbolic tangent function to the data near the phase transition points observed in Figures \ref{rg-epsilonlow}(a) and \ref{rg-epsilonlow}(b). For $\sigma_c \leq \sigma_m$, the critical $\epsilon_{mc}^*$ value increases monotonically with crowder size up to $\sigma_c = \sigma_m$. Similar to the size effect in depletion-mediated collapse, the smallest crowder size considered ($\sigma_c = 0.4$) induces the maximum compaction. Conversely, for $\sigma_c > \sigma_m$, an increase in crowder size correlates with a decrease in the critical $\epsilon_{mc}^*$, with a notable crossover in both $R_g$ and asphericity $\alpha$ values near $\epsilon_{mc} \approx 0.77$, corresponding to the critical $\epsilon_{mc}^*$ for $\sigma_c = \sigma_m = 1.0$. The phase diagram in the $\sigma_c-\epsilon_{mc}$ phase space (Figure \ref{phasediagramemmlow}) visually maps the boundary between the extended ($E$) and collapsed ($CB$) phases, based on this analysis of the explored parameter space.

\begin{figure}
\includegraphics[width=\columnwidth]{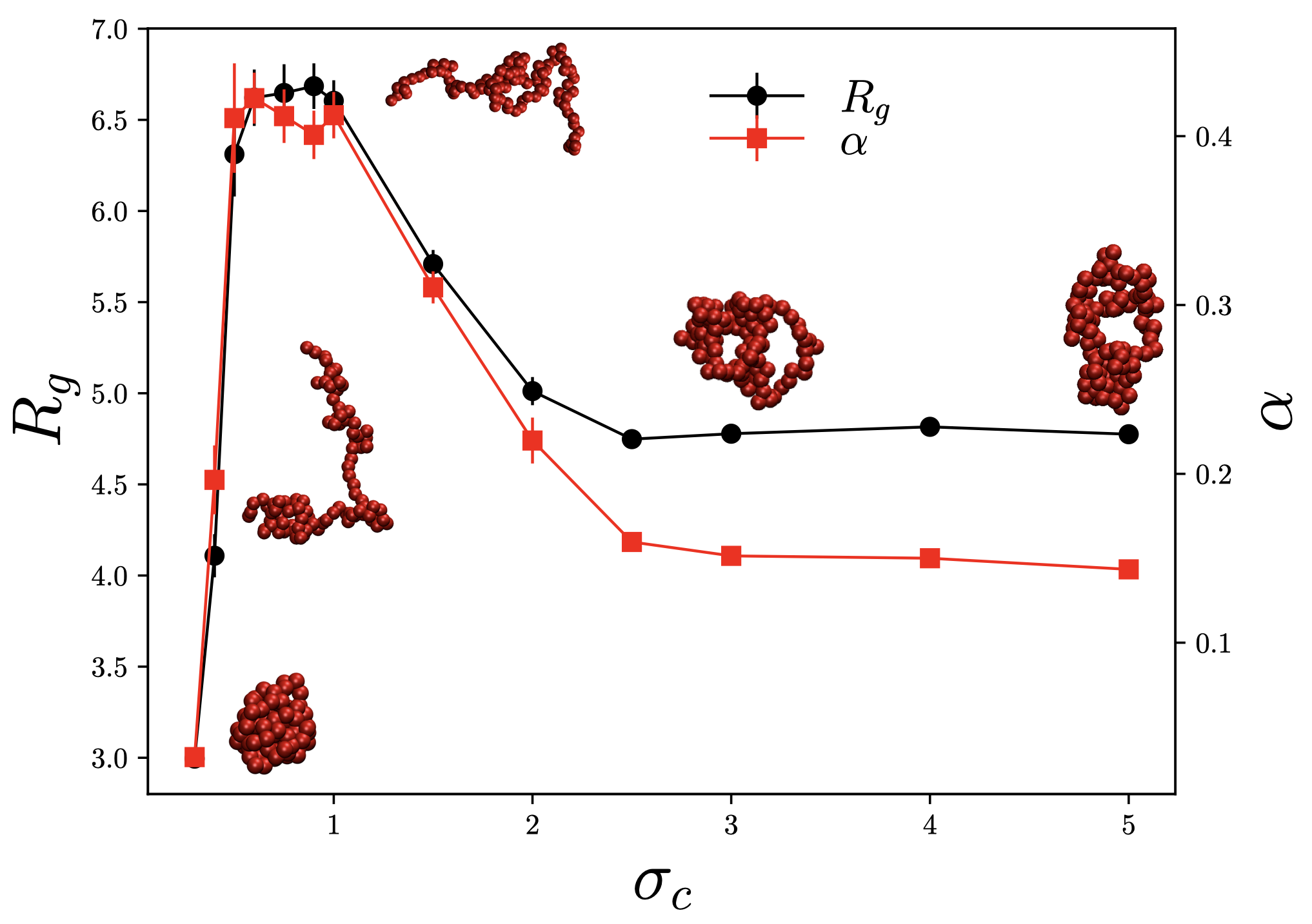}
 \caption{Collapse-extended-collapse ($CB$-$E$-$CB$) transitions as a function of crowder sizes for weakly self-attractive polymer ($\epsilon_{mm}=0.1$) for $\epsilon_{mc}=0.7$, near the transition point in $\sigma-\epsilon_{mc}$ phase space, measured in terms of $R_g$ and asphericity $\alpha$. Typical configurations of polymer along this transition are also shown.}
\label{Rgwscc}
\end{figure}

We now validate the derived phase boundary, which delineates the extended ($E$) and collapsed ($CB$) phases, based on the $R_g$ analysis. By examining a vertical cross-section of the phase diagram across different crowder sizes, particularly for $\epsilon_{mc}^* = 0.7$ near the transition point at $\sigma_c = \sigma_m$, we expect reentrant behavior to emerge if the phase boundary shape is accurate. Specifically, the polymer is predicted to exhibit a collapsed phase at very small $\sigma_c$ values, transition into an extended state around $\sigma_c \approx 1.0$, and then revert to a collapsed state for larger crowder sizes. The results of this analysis are shown in Figure \ref{Rgwscc}, where clear variations in the radius of gyration ($R_g$) and asphericity are observed as a function of crowder size. As the crowder size approaches $\sigma_c \approx 1.0$, both $R_g$ and asphericity increase, signaling a transition from a collapsed to an extended polymer conformation. This behavior aligns well with the $E$ phase region identified in the phase diagram, confirming the accuracy of the phase boundary shape. For crowder sizes larger than 1.0, the phase diagram reveals a sharp decrease in both $R_g$ and asphericity, indicating a reentry into the collapsed $CB$ phase. It is important to note, however, that the collapsed phases at the extremes of the crowder size spectrum exhibit distinct characteristics. These differences in polymer conformation between small and large crowder-induced collapsed states suggest underlying mechanistic variations in how crowder size influences polymer structure, which we discuss further below.

The delineation of the phase boundary in the $\sigma_c-\epsilon_{mc}$ phase diagram can be understood by examining how crowder size, relative to monomer size, influences the transition from the extended ($E$) phase to the collapsed ($CB$) phase. The critical interaction parameter, $\epsilon_{mc}^*$, required for this transition depends on the size of the crowders compared to the monomers. For crowders smaller than or comparable to the monomer size ($\sigma_c \leq \sigma_m$), a lower value of $\epsilon_{mc}^*$ is needed to induce the transition to the $CB$ phase. This is because smaller crowders can densely adsorb onto the polymer chain, covering more surface area. The increased surface coverage enhances the effective interaction between the polymer and crowders, thereby lowering the $\epsilon_{mc}^*$ threshold for the extended-to-collapsed phase transition via bridging interactions. Conversely, when crowders are larger than the monomers ($\sigma_c > \sigma_m$), they create interstitial spaces around the polymer, sequestering a significant number of monomers. This leads to effective bridging of polymer segments by the larger crowders, promoting the transition to the $CB$ phase at a lower $\epsilon_{mc}^*$ through a different mechanism. In both cases, crowder size plays a crucial role in determining the spatial arrangement and interaction dynamics between the polymer and crowders, directly influencing the interaction strength required for the phase transition. This twofold effect of crowder size aligns with earlier studies, which showed that large colloidal particles can adsorb polymers and induce collapse in self-repulsive polymers at high crowder densities~\cite{d2016phase,lafuma1991bridging,garcia2020polymer}. The overall shape of the phase boundary in the $\sigma_c-\epsilon_{mc}$ phase space is thus governed by the interplay between crowder size, polymer-crowder interaction strength, and the resulting adsorption patterns. Smaller crowders require lower $\epsilon_{mc}^*$ due to their ability to enhance surface interactions, while larger crowders promote effective monomer bridging at lower $\epsilon_{mc}^*$, albeit through a different mechanism.

\subsection{\label{sec:SubCollpasedphase} Sub phase in collapsed phase: confined collapse and bridging collapse}

While the polymer chain adopts a collapsed phase once the monomer-crowder attraction exceeds a certain threshold, the structure of the collapse and the underlying mechanisms differ significantly depending on crowder size. Though bridging interactions are the primary driving force, whether the crowders bridge polymer segments or the polymer bridges the crowders depends on crowder size and significantly impacts the local structure of the collapsed phase. When the crowder size is smaller than or comparable to the monomer size ($\sigma_c \leq \sigma_m$), crowders can penetrate the polymer globule, inducing a collapse from within. This is referred to as the bridging collapse (CB) phase. In contrast, when the crowders are larger than the monomers ($\sigma_c > \sigma_m$), the crowders cannot penetrate the polymer structure. Instead, the polymer is confined within the interstitial spaces between the larger crowders, leading to what we term the confined collapse (CC) phase, as illustrated in Figure \ref{varySigma}. Although both scenarios result in a collapsed state, the structural differences are notable. For instance, variations in the radius of gyration ($R_g$), as shown in Supplementary Figure S1(a), highlight these distinctions. In the CB phase, smaller crowders occupy the space between polymer segments, whereas in the CC phase, the polymer is confined within the voids between larger crowders. Despite these differences in structure, both phases exhibit collapsed behavior, as indicated by similar asphericity values (Supplementary Figure S1(b)).
\begin{figure}
\includegraphics[width=\columnwidth]{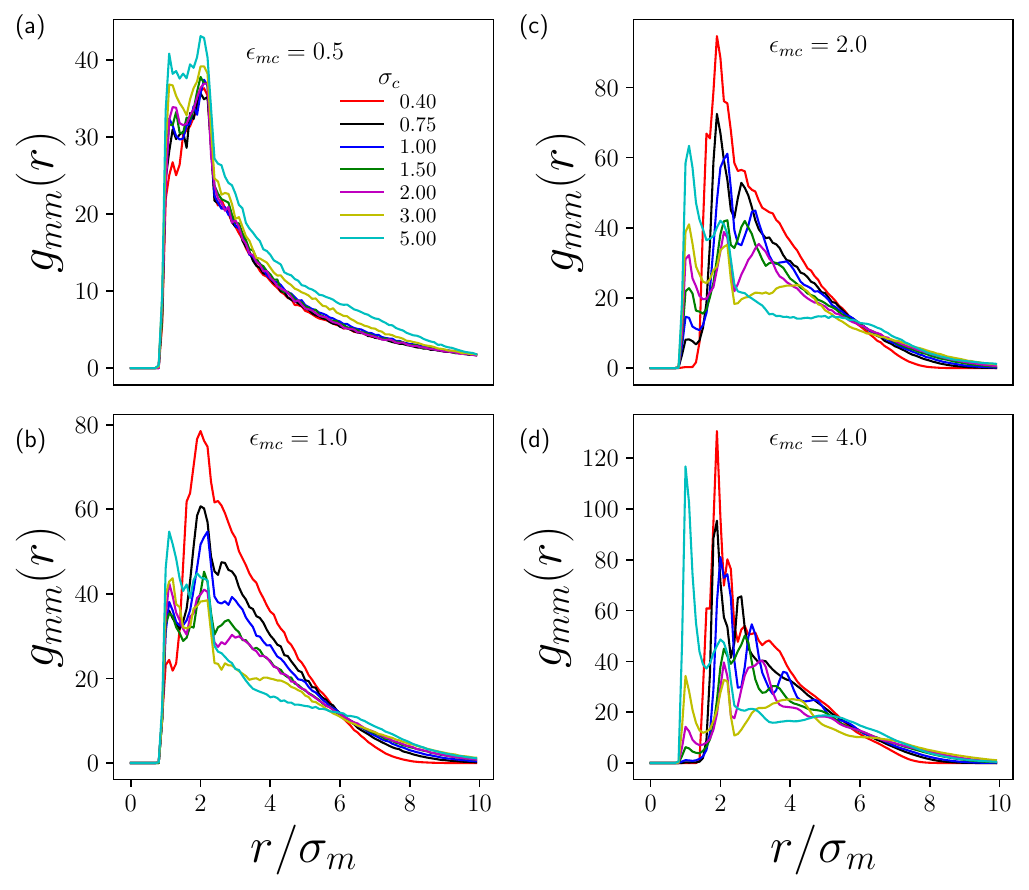}
 \caption{Radial distribution function between monomer-monomer $g_{mm}(r)$ for different values of crowder sizes for (a) $\epsilon_{mc}$=0.5 (b) $\epsilon_{mc}$=1.0 (c) $\epsilon_{mc}$=2.0 (d) $\epsilon_{mc}$=4.0 }
\label{gofrmm_emmlow}
\end{figure}

To further quantify these differences, we analyze the radial distribution function ($g_{mm}(r)$) between monomers at different values of $\epsilon_{mc}$. Before the transition (at $\epsilon_{mc}=0.5$), $g_{mm}(r)$ shows similar behavior across all crowder sizes. However, after the transition (at $\epsilon_{mc}=4.0$), the behavior diverges: for $\sigma_c \leq \sigma_m$, the first peak in $g_{mm}(r)$ is absent, suggesting that crowders are located between monomers, characteristic of the CB phase. Conversely, for $\sigma_c > \sigma_m$, the first peak remains, indicating that monomers are closely packed together, typical of the CC phase. To confirm that these are indeed distinct phases, we calculate the number of bridging crowders, $N_{bc}$, defined as the number of crowders in simultaneous contact with at least six monomers. The variation of $N_{bc}$ with crowder size is shown in Supplementary Figure S3. For larger crowder sizes, $N_{bc}$ increases, and the transition point is identified by fitting the $N_{bc}$ vs. $\sigma_c$ curve with a hyperbolic tangent function, pinpointing the transition at approximately $\sigma^*_c = 1.19$.

\begin{figure}
\includegraphics[width=\columnwidth]{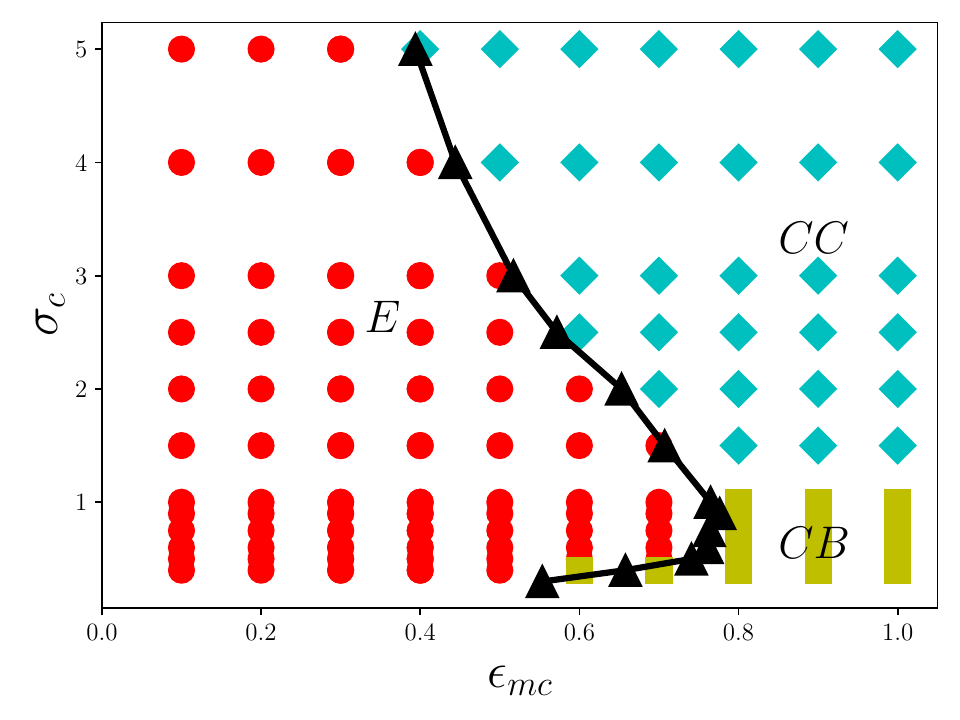}
 \caption{Phase diagram for polymer conformation in $\sigma_{c}$-$\epsilon_{mc}$ plane for weakly self-attractive neutral polymer, $\epsilon_{mm}=0.1$, obtained from the analysis of the chain structure in terms or radius of gyration $R_g$. The simulated systems corresponding to extended phase (E), bridging crowder induced collapsed phase (CB)  and bridging crowder induced confined collapse (CC) are colored in red, yellow and cyan respectively. The phase boundary is shown in black.}
\label{phasediagramemmlow1}
\end{figure}

This analysis confirms that for $\sigma_c \leq \sigma_m$, the polymer is in the bridging collapse (CB) phase, while for $\sigma_c > \sigma_m$, the polymer adopts the confined collapse (CC) phase. The updated phase diagram, showing the boundaries between these two subphases, is presented in Figure \ref{phasediagramemmlow1}. 

\subsection{\label{sec:highEmm}Collapse to collapse transition via bridging crowders}
\begin{figure}
\includegraphics[width=\columnwidth]{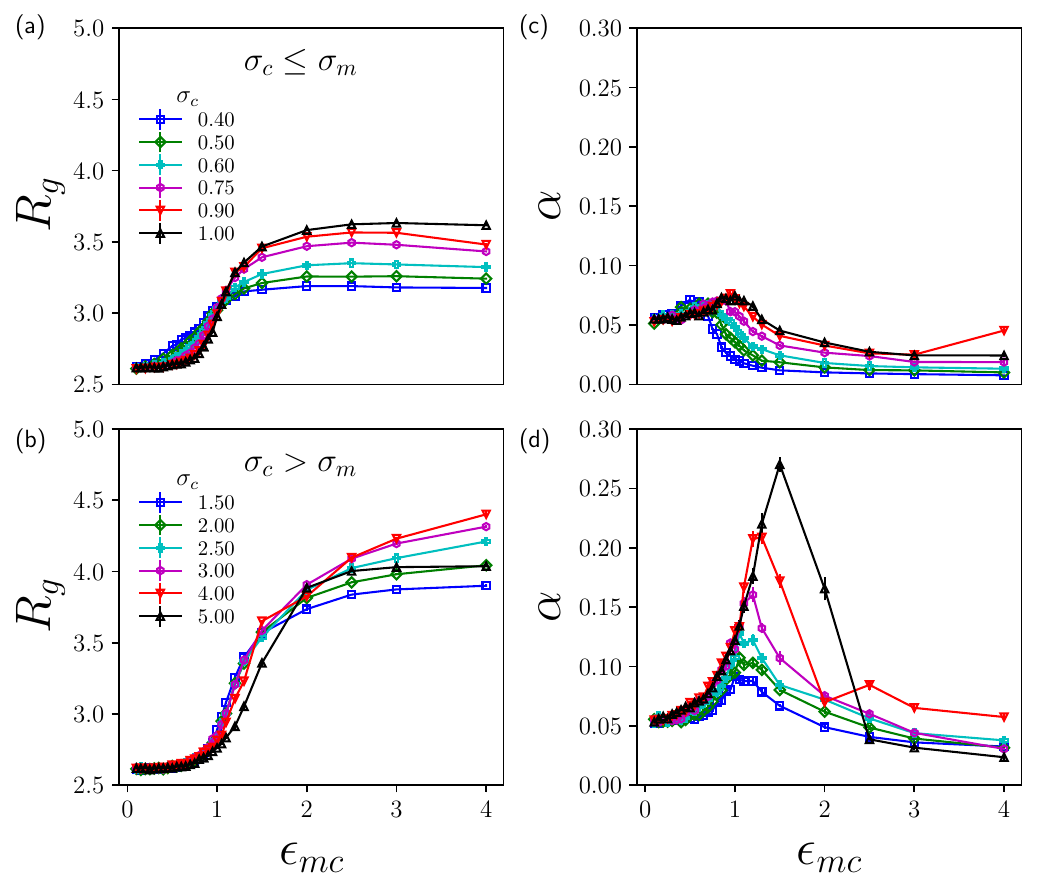}
 \caption{Variation of radius of gyration, $R_g$ (a, b), and asphericity, $\alpha$ (c, d), as a function of polymer-crowder attractive interaction strength, $\epsilon_{mc}$, for $\sigma_c \leq \sigma_m$ and $\sigma_c > \sigma_m$. The data are for strongly self-interacting polymer, $\epsilon_{mm}=1.0$. }
\label{rg-epsilonhigh}
\end{figure}

In this section, we explore the role of crowder sizes in the transition between two distinct collapsed polymer conformations: one induced by strong self-interactions (CI) and the other facilitated by bridging due to attractive crowder interactions (CB). The variation in the radius of gyration, $R_g$, with respect to the monomer-crowder interaction parameter, $\epsilon_{mc}$, is shown for crowders smaller and larger than the monomer size in Figures \ref{rg-epsilonhigh}(a) and \ref{rg-epsilonhigh}(b), respectively. At a fixed self-interaction strength ($\epsilon_{mm} = 1.0$), the polymer adopts a collapsed state (CI) at low $\epsilon_{mc}$ across all crowder sizes, driven primarily by intrapolymer attractions. This CI phase persists even in the absence of crowders.

\begin{figure}
\includegraphics[width=0.8\columnwidth]{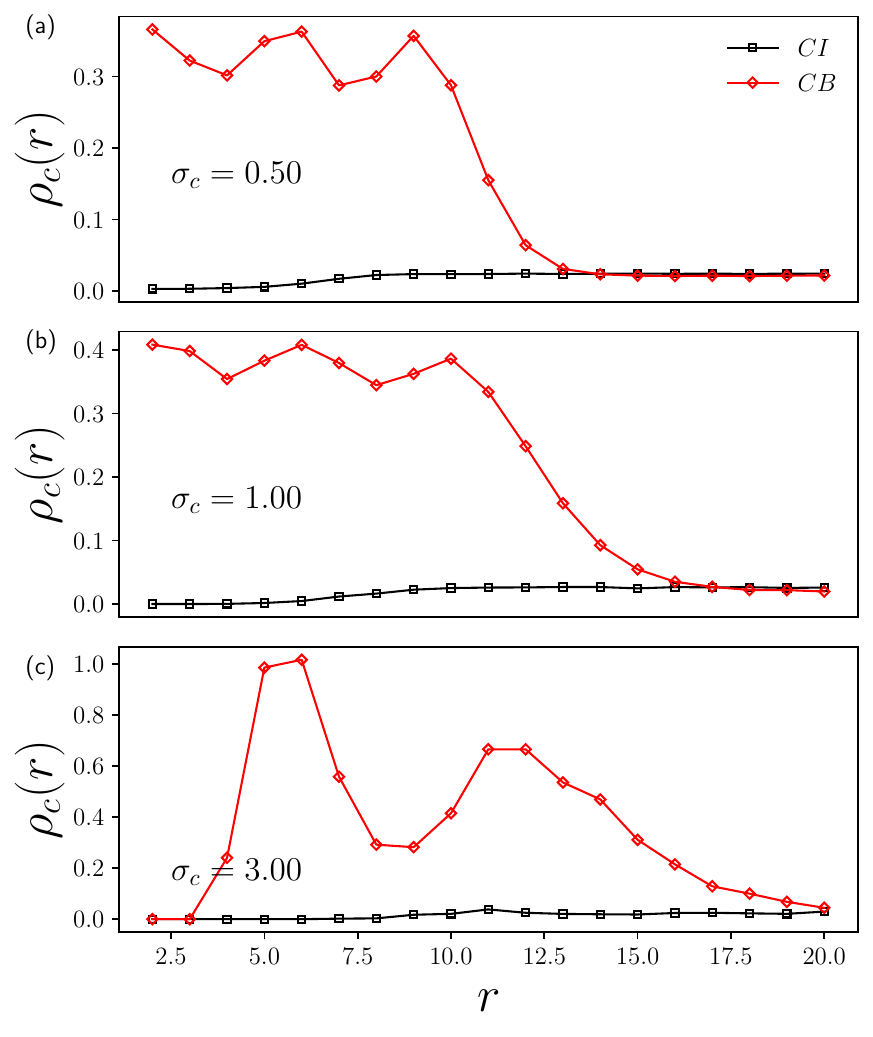}
 \caption{The radial density function of crowders, $\rho_c(r)$, measured from the center of mass of the collapsed neutral polymer for different crowder sizes, distinguishes between the $CI$ phase ($\epsilon_{mc} = 0.1$) and the $CB$ phase ($\epsilon_{mc} = 4.0$).}
\label{CI_CB}
\end{figure}

As $\epsilon_{mc}$ increases, the system transitions from the CI phase to a bridging-induced collapsed phase (CB) across all crowder sizes. While the $R_g$ values follow similar trends for both small and large crowders, the asphericity profiles show significant divergence. For smaller crowders, asphericity remains consistently low during the transition from the intrapolymer-driven collapsed phase (CI) to the crowder-mediated collapsed phase (CB), as shown in Figure \ref{rg-epsilonhigh}(c). This sustained low asphericity suggests a continuous collapsed state, unaffected by increasing $\epsilon_{mc}$. In contrast, for larger crowders, the transition is marked by an intermediate extended phase at certain $\epsilon_{mc}$ values, as shown in Figure \ref{rg-epsilonhigh}(d), indicating a more complex interaction dynamic. This difference arises from the different ways small and large crowders are accommodated within the collapsed polymer structure. Smaller crowders integrate seamlessly among the monomers, maintaining the collapsed conformation with minimal structural adjustment. On the other hand, incorporating larger crowders requires substantial conformational rearrangements due to strong intrapolymer attractions, which increases the energetic cost. Once $\epsilon_{mc}$ exceeds the intrapolymer attraction threshold ($\epsilon_{mm}$), crowder interactions dominate, prompting a structural transformation. Further increases in $\epsilon_{mc}$ lead to the reformation of a collapsed structure, where the polymer becomes embedded in the interstitial spaces formed by large, surface-bound crowders, similar to behavior seen in weakly self-attracting polymers.

The distinction between the CI and CB phases, particularly for intermediate-sized crowders ($\sigma_c \leq 3$), is further clarified through the analysis of the radial density distribution of crowders around the polymer’s center, $\rho_c(r)$. This metric provides insight into the spatial distribution of crowders relative to the polymer globule in both phases. In both small and large crowder cases, the CI phase is characterized by an exclusion zone near the center of the globule, while the CB phase shows a significant number of crowders within the globule. This difference in crowder distribution, as illustrated in Figure \ref{CI_CB}, highlights the distinct natures of the CI and CB phases. Additionally, for very large crowders ($\sigma_c > 3$), no crowders are found inside the collapsed conformations, similar to the behavior observed in weakly self-attracting polymers.

\begin{figure}
\includegraphics[width=\columnwidth]{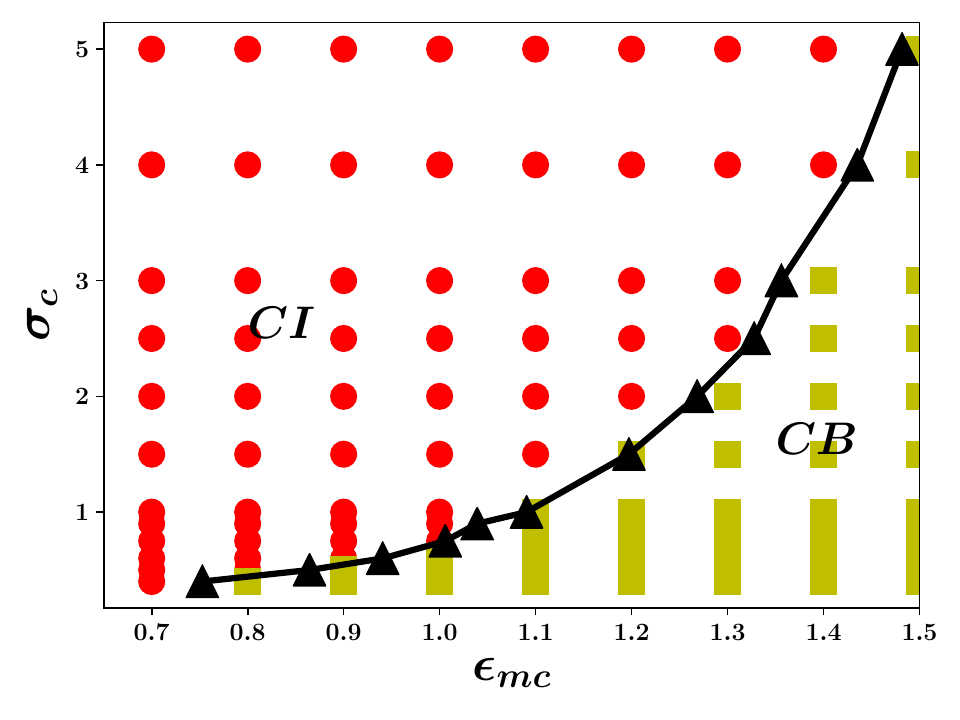}
 \caption{Phase diagram for polymer conformation in $\sigma_{c}$-$\epsilon_{mc}$ plane for strongly self-attractive neutral polymer, $\epsilon_{mm}=1.0$, obtained from the analysis of the chain structure in terms or radius of gyration $R_g$. The simulated systems corresponding to strong intra attraction induced collapse (CI) and bridging crowder induced collapsed phase (CB) are colored in red and yellow respectively. The phase boundary is shown in black.}
\label{phasediagramemmhigh}
\end{figure}

To construct the phase diagram showing the transition from the CI phase (dominated by intrapolymer attractions) to the CB phase (induced by crowder bridging), as a function of the monomer-crowder interaction parameter ($\epsilon_{mc}$) and crowder size ($\sigma_c$), we followed a methodology similar to that used in the previous section. The transition points ($\epsilon_{mc}^*$) were calculated based on changes in the radius of gyration ($R_g$). The resulting phase diagram is shown in Figure \ref{phasediagramemmhigh}. The shape of the phase boundary can be understood by considering the effect of crowder size on polymer transition dynamics. As crowder size increases, a higher $\epsilon_{mc}$ is required to facilitate the transition from the CI to the CB phase. This is due to the increased energy cost of altering the polymer’s conformation to accommodate the larger crowders. As a result, the critical interaction parameter $\epsilon_{mc}^*$ for the CI to CB transition increases monotonically with crowder size. The larger the crowders, the more substantial the conformational rearrangement required for the polymer to transition into the crowder-bridging-induced collapse, leading to a higher $\epsilon_{mc}$ threshold. The bridging-induced collapsed phase is identical for both weakly and strongly self-attractive polymers, as demonstrated by the overlap of crowder densities within the collapsed phases across different crowder sizes, as shown in Supplementary Figure S4.

\begin{figure}
\includegraphics[width=\columnwidth]{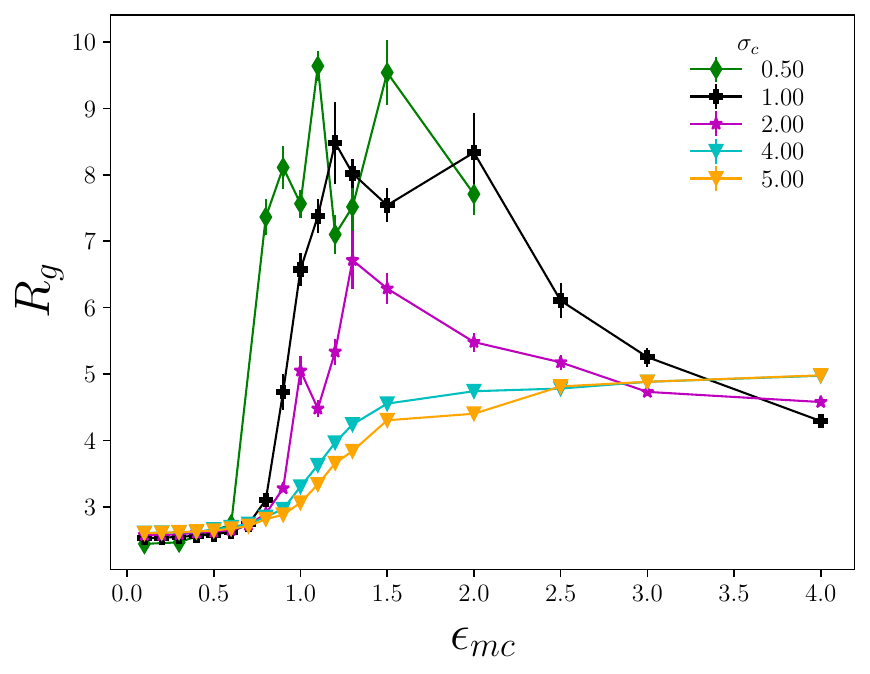}
 \caption{The variation of $R_g$ with $\epsilon_{mc}$ for a strongly self-interacting neutral polymer, $\epsilon_{mm}=1.0$, at high crowder density $\phi_c=0.33$}
\label{Rghighdensity}
\end{figure}

Previous studies~\cite{huang2021chain,garg2023conformational} have documented an intriguing phenomenon where, at high crowder densities (specifically when $\sigma_c = \sigma_m = 1.0$), a polymer in the collapsed phase (CI) transitions to a fully extended state before reverting to another collapsed phase (CB). In this section, we investigate whether similar reentrant behavior occurs for crowder sizes smaller or larger than the monomer size. Notably, this reentrant behavior was not observed at the lower crowder density ($\phi_c = 0.025$) used in previous studies, as shown in Figure \ref{rg-epsilonhigh}. To explore reentrant behavior across a broader range of crowder sizes, we conducted simulations at a significantly higher crowder density ($\phi_c = 0.33$), varying the monomer-crowder interaction parameter ($\epsilon_{mc}$) accordingly. The results, presented in Figure \ref{Rghighdensity}, reveal a monotonic transition from a collapsed state to an extended state as crowder size increases up to $\sigma_c = 4.0$. Beyond this size, no transition to an extended state occurs. This shift in polymer conformation suggests a change in effective solvent conditions, moving from poor (favoring collapse) to good (favoring extension) and back to poor, demonstrating reentrant behavior. The absence of this behavior for larger crowders at the studied density and $\epsilon_{mc}$ range suggests a connection between the overall polymer size and crowder size. Specifically, when the polymer’s size in its collapsed phase matches the interstitial volume between crowders, the crowders are unable to induce a fully extended state, preventing reentrant behavior. The collapse induced by purely repulsive crowders is attributed to entropic effects, which stabilize the collapsed state by maximizing the available space for crowders, leaving no crowders inside the collapsed structure. Conversely, large attractive crowders induce similar collapsed conformations by confining the polymer within the interstitial spaces between them. Interestingly, although both types of collapse—induced by repulsive and large attractive crowders—exclude crowders from the polymer's interior, the resulting conformations differ. Depletion-induced collapse tends to produce more spherical or symmetrical shapes, while collapse driven by large attractive crowders forms "finger-like" configurations that optimize interactions with surrounding crowders. This difference highlights the complex interplay between polymer conformation, crowder size, and interaction type, emphasizing the diverse ways in which crowders influence polymer behavior in crowded environments.

\section{\label{sec:summary} Discussion}
In this study, we examined the influence of crowders on polymer conformation, inspired by biological contexts where biopolymers exist in crowded environments. This underscores the importance of understanding how both synthetic and biological crowders modulate the stability of macromolecules through diverse interactions~\cite{zimmerman1991estimation,ellis2003join,ellis2001macromolecular,zimmerman1993macromolecular,zhou2008macromolecular,ellis2001macromolecular1,foffi2013macromolecular,nayar2020small,zangi2009urea,mardoum2018crowding,nakano2017model,shin2015kinetics}. The traditional view of crowders as steric or volume-exclusion entities is fundamental for understanding depletion-induced macromolecular collapse, which is driven by the entropic stabilization of collapsed conformations~\cite{minton1981excluded,zhou2008macromolecular,asakura1958interaction,bhat1992steric,minton1981effect}. Prior studies have demonstrated that crowder size can significantly influence the extended-to-collapse phase behavior of polymers~\cite{kang2015effects,liu2020non,kang2015effects,sharp2015analysis,kim2015polymer,chen2015polymer}. In crowded cellular environments, a combination of soft attractive and hard volume-exclusion interactions plays a crucial role in determining biomacromolecular conformations and interactions~\cite{sarkar2013soft,sagle2009investigating,zangi2009urea,street2006molecular}. Recent studies indicate that attractive interactions between polymers and crowders can promote coil-globule transitions, stabilizing collapsed or folded conformations~\cite{antypov2008computer,heyda2013rationalizing,rodriguez2015mechanism,sagle2009investigating,huang2021chain,ryu2021bridging,brackley2020polymer,brackley2013nonspecific,barbieri2012complexity,garg2023conformational}. This transition becomes particularly pronounced when polymer-crowder attractions exceed a critical threshold, facilitating "bridging interactions" that induce polymer collapse.

Our molecular dynamics simulations extend this understanding by focusing on the role of crowder size in the conformational phase diagram of neutral polymers. A key finding is that crowder size significantly affects the extended-to-collapse phase behavior, confirming and building upon prior research on polymers in crowded environments. Both soft attractive and hard volume-exclusion interactions between polymers and crowders are essential in driving these transitions. When polymer-crowder attraction exceeds a critical threshold, bridging interactions stabilize the collapsed conformation. By mapping the conformational phase diagram of neutral polymers in the $\sigma_c$ and $\epsilon_{mc}$ plane for both weakly and strongly self-interacting polymers, we identified three distinct phases: the collapsed intrapolymer attraction-driven phase (CI), the extended phase (E), and the crowder-bridging-induced collapsed phase (CB). In the CI phase, strong intrapolymer attractions dominate, with no crowders within the polymer structure. The E phase represents the extended polymer conformation, while the CB phase is a collapsed state induced by bridging interactions with crowders, where a significant number of crowders are embedded within the collapsed structure. Interestingly, the bridging-induced collapse phases for both weakly and strongly self-interacting polymers are identical, underscoring the diminished role of intrapolymer interactions in the presence of strong polymer-crowder attraction.

For weak intrapolymer interactions ($\epsilon_{mm}=0.1$), we observed two distinct phases: an extended phase (E) at low polymer-crowder interaction values, and a collapsed phase (CB) resulting from bridging at higher $\epsilon_{mc}$ values. Notably, the critical $\epsilon_{mc}$ value required for collapse decreases as crowder size decreases when crowders are smaller than the monomers. Conversely, for larger crowders, the transition to the CB phase occurs at higher $\epsilon_{mc}$ values. Additionally, the mechanisms driving collapse differ based on crowder size, leading to two distinct types of collapse: (1) CB Phase (Bridging Collapse): Smaller crowders integrate within the polymer structure, bridging monomers together into a compact form. (2) CC Phase (Confining Collapse): Larger crowders act as external confinements, forming spaces around the polymer, inducing collapse without integrating into the polymer structure. These differences lead to distinct morphologies, such as finger-like extensions or looped conformations, with very large crowders potentially inducing a transition to two-dimensional conformations.

In the case of strong intrapolymer interactions ($\epsilon_{mm}=1.0$), the polymer adopts a collapsed state (CI) at low $\epsilon_{mc}$ across all crowder sizes, driven by intrapolymer attraction. As $\epsilon_{mc}$ increases, the polymer transitions from the CI phase to a bridging-induced CB phase. The critical value of $\epsilon_{mc}$ required for this transition increases with crowder size: smaller crowders induce the transition at lower $\epsilon_{mc}$ values, while larger crowders require higher $\epsilon_{mc}$. For larger crowders, we observed a partial extension phase, indicated by increasing asphericity around $\epsilon_{mc} \approx 1.0$, before the polymer returns to a fully collapsed state at higher $\epsilon_{mc}$ values. In contrast, for smaller crowders, asphericity remains low throughout the transitions, reflecting a consistently collapsed state. We further verified these phase distinctions by calculating the number of crowders incorporated into the collapsed structure for different crowder sizes. Finally, we investigated reentrant behavior at high crowder densities, as reported in previous studies~\cite{huang2021chain,garg2023conformational}. Our results show that smaller crowders enhance reentrant behavior, while larger crowders suppress it. For very large crowders ($\sigma_c > 4\sigma_m$), reentrant behavior was absent. Experimentally, the expansion and compaction of double-stranded DNA in crowded environments aligns with our simulation results, where smaller crowders induce more pronounced effects~\cite{gupta2017compaction}.

Several future research directions remain. In biological systems, crowders vary in size and shape, which may influence polymer conformation differently than the monodisperse crowders used in this study. Previous work has shown that polydisperse crowders can modulate polymer compaction and phase transitions~\cite{rivas2004life,zhou2008macromolecular,cheung2005molecular,kim2015polymer}. Additionally, dynamic crowders—those that change size, shape, or interaction properties over time—may have complex effects on polymer stability~\cite{minton2006can,das2020shape}. Another promising avenue is the role of active crowders in biological systems, where their activity could further influence polymer behavior~\cite{yan2023conformation,yan2024attractive,tan2021effects,feng2021tunable}.

\section*{Acknowledgment}
 We thank the HPC facility at the Institute of Mathematical Sciences for providing computing time. We are grateful for the insightful and stimulating discussions with R. Rajesh, which critically contributed to the development of this work.

\bibliography{neutral}
\end{document}